\documentstyle[12pt]{article}

\newcommand{\be}{\begin{equation}}
\newcommand{\ee}{\end{equation}}
\newcommand{\bea}{\begin{array}}
\newcommand{\ea}{\end{array}}
\newcommand{\beqa}{\begin{eqnarray}}
\newcommand{\eeqa}{\end{eqnarray}}
\newcommand{\bean}{\begin{eqnarray*}}
\newcommand{\eean}{\end{eqnarray*}}
\newcommand{\eqn}[1]{(\ref{#1})}

\def\up#1{\leavevmode \raise.16ex\hbox{#1}}

\def\sqr#1#2{{\vcenter{\vbox{\hrule height.#2pt
        \hbox{\vrule width.#2pt height#1pt \kern#1pt
          \vrule width.#2pt}
        \hrule height.#2pt}}}}

\newcommand{\gapproxeq}{\lower .7ex\hbox{$\;\stackrel{\textstyle >}{\sim}\;$}}
\newcommand{\lapproxeq}{\lower .7ex\hbox{$\;\stackrel{\textstyle <}{\sim}\;$}}

\def\thebibliography#1{{\bf REFERENCES\markboth
 {REFERENCES}{REFERENCES}}\list
 {[\arabic{enumi}]}{\settowidth\labelwidth{[#1]}\leftmargin\labelwidth
 \advance\leftmargin\labelsep
 \usecounter{enumi}}
 \def\newblock{\hskip .11em plus .33em minus -.07em}
 \sloppy
 \sfcode`\.=1000\relax}

\begin{document}
\begin{flushright}
DSF 12/98\\ UAHEP 982\\ gr-qc/9805022
\end{flushright}

\begin{center}
{\bf \large{Comments on the Non-Commutative Description of\\ Classical
Gravity }}\\ {G. Bimonte$^{a}$, R. Musto$^{a}$, A. Stern$^{b}$ and P.
Vitale $^{a}$}
\end{center}
\vspace{1cm}

\begin{center}
{\it a)  Dipartimento di Scienze Fisiche, Universit\`a di Napoli Federico
II,\\ Mostra d'Oltremare, Pad.19, I-80125, Napoli, Italy; \\ INFN, Sezione
di Napoli, Napoli, ITALY.\\
\small e-mail: \tt bimonte,musto,vitale@napoli.infn.it } \\
{\it b) Department of Physics, University of Alabama,\\
Tuscaloosa, AL 35487, USA.\\
\small e-mail: \tt astern@ua1vm.ua.edu }\\
\end{center}

\begin{abstract}
We find a one-parameter family of Lagrangian descriptions for classical
general relativity in terms of tetrads which are not c-numbers.  Rather, they
obey  exotic commutation relations.  These noncommutative properties drop out
in the metric sector of the theory, where the Christoffel symbols and
 the Riemann tensor are ordinary commuting objects and they are given by the
usual expression in terms of the metric tensor.
Although the metric tensor is not a c-number, we
argue  that all measurements one can make
 in this theory are associated with c-numbers, and
thus   that  the common invariant sector
of our one--parameter family of deformed gauge theories (for the case of zero
 torsion) is physically equivalent to Einstein's general relativity.

\end{abstract}

It is well known that $3+1$ gravity admits a gauge theory
description\cite{UKH}. In this description, the connection one forms
correspond to the tetrads and spin connections, while the dynamics is given
by the Palatini action.  The   gauge group is the  Poincar\'e group, although
the action  is only invariant under local Lorentz transformations.

In a couple of recent papers \cite{noi2+1, noi3+1} we obtained a
generalization of the gauge theory description of general relativity where
the  gauge group is replaced by a q-gauge group\cite{cast1}. This result
was first achieved in three space-time dimensions \cite{noi2+1} using a
deformed Chern--Simons action\cite{noiPL} and the quantum Poincar\'e group
$ISO_q(2,1)$ \cite{cast2}.  In the resulting description of 2+1 gravity,
the dreibeins and spin--connections are not c-numbers, but instead, obey
nontrivial braiding relations. In four space-time dimensions\cite{noi3+1},
one  uses a deformed Palatini action and the quantum Poincar\'e group
$ISO_q(3,1)$ \cite{cast96}. Once again, the connections obey nontrivial
braiding relations. There is, in fact, a one-parameter family of such
theories, parametrized by $q$, and the usual (undeformed) theory is
obtained in the limit $q\rightarrow 1$. The metric tensor, which is needed
for recovering Einstein's theory, can be constructed as a suitable bilinear
in the tetrads, and it reduces to the usual expression when  $q\rightarrow
1$.
  A remarkable feature of our description  (in the
absence of torsion) is that the entire one--parameter family of
descriptions (including the $q=1$ case) has a common metric sector. That
is,   {\it
all descriptions lead to the same classical dynamics - that given by the
Einstein equations}.  In this regard,  the components of
the metric tensor $g_{\mu\nu}$ commute among themselves and the field
equations they satisfy are formally identical to those of the ordinary
theory associated with $q=1$.  Remarkably, the relevant fields of the
metric theory, such as Christoffel symbols and the Riemann tensor, turn out
to be ordinary commuting fields given by the usual expressions in term of
$g_{\mu\nu}$.

In this letter, after giving a concise review of
 our formulation of gravity,
we shall present new results concerning two important
issues that were not addressed in our previous papers.  First, we
 shall show that
for $q$ real or equal to a phase there exists a set of reality conditions
for our non-commuting tetrads and spin-connections which are consistent
with their commutation properties, and ensure that the space-time metric is
real. Second, we shall argue  that the equivalence of our deformed
gauge theory with classical general relativity goes beyond  the formal
arguments sketched above.  By `formal', we are alluding to the fact that
our deformed metric tensor does not commute with all the connections of the
theory and hence is not a c-number.  One may therefore question its physical
relevance, since the metric is necessary for defining distances. Since  the
Einstein-Hilbert action ${\cal S}_{EH}$ contains   $g_{\mu\nu}$, one might
 conclude that  ${\cal S}_{EH}$ too is  not a c-number, causing new obstacles,
  say if one were to
attempt a path integral quantization. We will show below that despite the
fact that the metric is noncommuting, the ratio of $any$ two quantities
with the units  of  `length' can be made to be in the center of the
algebra, and hence any actual distance measurement
 one makes in this theory can be associated with
a c-number.  Moreover, once we allow dimensionful quantities (say, for
example, the analogue of Newton's constant) to have non trivial commutation
properties, we can argue that {\it all} possible dimensionless ratios, and
hence {\it all} possible measurements  are associated with c-numbers, and
hence are physically meaningful. Concerning the action integral, although
it is not a c-number, we find that it has definite commutation properties.
That is, all terms appearing in the integrand have the same commutation
relations. (We should point out that this, in general, is not guaranteed
when dealing with theories of noncommuting fields.)   ${\cal S}_{EH}$
 can then be made to be central in the algebra of fields, when one
allows for the dimensionful coefficient,
 here the analogue of Newton's constant, to be
noncommuting. Its commutation properties are precisely those which are
needed to make dimensionless ratios   into c-numbers.

It should be stressed  that the equivalence with the usual metric theory
holds, not only  for  pure gravity, but  also in the presence of matter,
provided there are no sources for torsion.  On the other hand,  the
different theories can be distinguished from one another, a) at the
classical level  when a non--zero torsion exists, each theory coupling to a
different kind of ``exotic'' matter. In addition, most certainly  the
different theories can be distinguished from one another, b) at the quantum
level.
 Upon applying the canonical formalism to the above system in any number of
dimensions, one  obtains a one--parameter family of  Hamiltonian
descriptions for  gravity.  The one-parameter family of canonical
formalisms is associated with inequivalent theories. This feature should
become significant when canonically quantizing such a  theory.
  We will not be concerned with a) or b)  in this
letter, and instead refer the
reader to \cite{noi3+1} for a detailed discussion of these and other
issues.

The starting point of the construction is a deformation $ISO_q(3,1)$ of the
Poincar\'e group. It can be described very simply in terms of a Lorentz
vector $z_a$ and a Lorentz matrix $\ell_{ab}$, whose elements, instead of
being c-numbers, have the following non-trivial commutation relations
\be
z^a ~{\ell_c}^{b} = q^{\Delta(b)} ~ {\ell_c}^{b} ~z^a ~,\label{4dcz}
\ee
where
$
\Delta(1)=-1 ~,\quad \Delta(2)=\Delta(3)=0 ~,\quad \Delta(4)=1 ~,
$ while all other commutation relations are trivial. The Lorentz metric
tensor is taken to be the following off-diagonal matrix:
\be
\eta=\pmatrix{ & & & 1\cr & 1 & & \cr & & 1 & \cr 1 & & & \cr}~.
\label{eta}
\ee
Then the  commutation relations \eqn{4dcz}  are consistent with the
Lorentz constraints \be
\ell_{ab}{ \ell_c}^b=
\ell_{ba} {\ell^b}_c= \eta_{ac}~, ~~
 \det \parallel \ell_{ab} \parallel =    1~,
\label {dolm}
\ee
 due to the identity
$
\eta_{ab}= q^{\Delta(a)+\Delta(b)} \eta_{ab} ~.$
$ISO_q(3,1)$ thus contains the undeformed Lorentz group.

We need now to formulate the reality properties of the group elements. Thus we
introduce a $*$-involution on the $q$--Poincar\'e group with the usual
property for the conjugation of a
 product, $(\alpha\beta)^*=
\beta^*\alpha^*$.  We shall demand that the commutation relations
(\ref{4dcz}) be preserved under conjugation, with $z^a$  real (i.e.,
$(z^a)^*=z^a$) and  $ {\ell_a}^b$ defining the Lorentz group.  We find that
this is possible for two cases:

\noindent
i) $q$ is a phase. Then the commutation relations (\ref{4dcz}) are
preserved with $(z^a)^*=z^a$ and  $ ({\ell_a}^b)^*={\ell_a}^b$.

\noindent
ii) $q$ is real.     The commutation relations (\ref{4dcz}) are preserved
with $(z^a)^*=z^a$ and  $ ({\ell_a}^b)^*$ proportional to
$\ell_{ab}={\ell_a}^d\eta_{db}$. It remains to ensure that the matrices
${\ell_a}^b$
 still define the Lorentz group.
For this we can choose the proportionality constants according to:
\be  ({\ell_a}^1)^*= {\ell_a}^4\;, \quad
 ({\ell_a}^2)^*= {\ell_a}^2\;, \quad
 ({\ell_a}^3)^*=- {\ell_a}^3\;, \quad
 ({\ell_a}^4)^*= {\ell_a}^1\;. \ee
 Although ${\ell_a}^b$ are not real, real matrices
$\tilde{\ell_a}^b$ can be constructed according to
\begin{eqnarray}
\tilde{\ell_a}^1& =&\frac1{\sqrt{2}}({\ell_a}^2 +i{\ell_a}^3)\;,\;\;\;
\tilde{\ell_a}^2 =\frac1{\sqrt{2}}({\ell_a}^1 +{\ell_a}^4) \cr
\tilde{\ell_a}^3& =&\frac{i}{\sqrt{2}}({\ell_a}^1 -{\ell_a}^4)\;,\;\;\;\;
\tilde{\ell_a}^4 =\frac1{\sqrt{2}}({\ell_a}^2 -i{\ell_a}^3) \;,
\end{eqnarray}
and it can be checked that these matrices satisfy the Lorentz conditions
(\ref {dolm}), e.g. $ {{\tilde\ell}_{a b}}{{\tilde
\ell}_c}^{\;\;b}={{\tilde\ell}_{b a }}{{\tilde
\ell}_{\;\;c}}^{b}= \eta_{ac}~.$

A set of  (left-)invariant Maurer-Cartan forms on $ISO_q(3,1)$ can
be constructed according to:
\be
\omega^{ab}=(\ell^{-1} d \ell)^{ab}\;\;,\;\;\;e^a=(\ell^{-1}dz)^a\;\;.\label{mc}
\ee
(Notice that the Lorentz constraints \eqn{dolm} imply that
$(\ell^{-1})_a^{\;\;b}=\ell^b_{\;a}$). The differential $d$, in these
formulae, satisfies all the usual properties of the undeformed case, like
$d^2=0$ and the Leibnitz rule. See ref. \cite{noi3+1}.
The commutation
properties of (the components of) the Maurer-Cartan one-forms follow easily
from Eqs.\eqn{4dcz}:
\beqa
\omega _\mu ^{ab}\omega _\nu ^{cd}&=&\omega _\nu ^{cd}\omega _\mu
^{ab}~,\cr
e_\mu ^a\omega _\nu ^{bc}&=&q^{\Delta (b)+\Delta (c)}~
\omega _\nu ^{bc}e_\mu^a~,\cr
e_\mu ^ae_\nu ^b&=&
q^{\Delta (b)-\Delta (a)}~ e_\nu ^be_\mu ^a~.
\label{4dcrstc}
\eeqa
As for their reality properties,  for i) $q$ equal to a phase, $e^a$
and $\omega^{ab}$ are real, while for ii) $q$ real, one finds:
\be
\omega^{ab*}=\omega_{ab}\;\;,\;\;\;e^{a*}=q^{-\Delta(a)}e_a\;.
\ee
Under infinitesimal (right) Poincar\'e transformations, the transformation
law of the Maurer-Cartan forms has the standard form:
\begin{eqnarray}
\delta \omega^{ab} & =& d\tau^{ab} + {\omega^a}_c \;\tau^{cb} -
{\omega^b}_c \;\tau^{ca} \;,\cr
\delta e^c & = & d\rho^c +
{\omega^{c}}_b \;\rho^b - \tau^{cb}\;e_b\;.  \label{4dtras}
\end{eqnarray}
but consistency with Eqs.\eqn{4dcrstc} implies that the gauge parameters
$\rho^a$ and $\tau^{ab}$ must be non-commuting elements such that:
$$
\;\;\rho^a \;\omega^{bc} = q^{\Delta(b)+\Delta(c) } \;\omega^{bc}\;
\rho^a\;,\;\;\;\;\;\;\;
\rho^a\; e^{b} = q^{\Delta(b)-\Delta(a)} \; e^{b}\;  \rho^a
$$
\be
\tau^{ab} \; e^c  = q^{-\Delta(a)-\Delta(b)} \; e^{c} \;
\tau^{ab}\;,\;\;\;\;\;\;\;~
\tau^ {ab} \omega^{cd} = \omega^{cd}  \tau^ {ab} \;. \label{crgpc}
\ee
When $e^a$ and $\omega^{ab}$ are given by (\ref{mc}), they
 satisfy a set of Maurer-Cartan equations ${\cal
R}^{ab}={\cal T}^a=0$, where the curvature ${\cal R}^{ab}$ and the torsion
and ${\cal T}^a $ have the usual expression
\begin{eqnarray}
{\cal R}^{ab} & =& d\omega^{ab}+ {\omega^a}_c\wedge \omega^{cb}\;,\nonumber
\\ {\cal T}^a & =& de^a+ {\omega^a}_b\wedge e^b\;.  \label{4dRT}
\end{eqnarray}
Even though Eqs.\eqn{mc}, \eqn{4dtras}, and \eqn{4dRT} all look identical
to the standard expressions, one should keep in mind that they involve
non-commuting quantities and so the ordering is crucial in all of them.

The passage to gauge theory is now achieved upon relaxing the flatness
conditions on the forms $e^a$ and $\omega^{ab}$, Eq.\eqn{mc}, while keeping
Eqs.(\ref{4dcrstc}-\ref{4dRT}), and pulling them back from the quantum
group to space-time. One thus ends up having a non-commuting set of tetrad
and spin-connection one-forms defined on space-time. Next one writes down a
locally Lorentz invariant action:
\be
{\cal S} =\frac{1}{4} \int_M \; q^{-\Delta(d)} \epsilon_{abcd} {\cal
R}^{ab}\wedge e^c \wedge e^d
\;,
\label{4dact}
\ee
$M$ is a four manifold and $\epsilon_{abcd} $ is the ordinary, totally
antisymmetric tensor with $\epsilon_{1234} =1$. The expression
(\ref{4dact}) differs from that of  the undeformed case by the
$q^{-\Delta(d)} $ factor. Note that this factor can be written differently
using the identity
\be
q^{\Delta(a)+\Delta(b)+ \Delta(c)+ \Delta(d)} \; \epsilon_{abcd}
=\epsilon_{abcd}\;\;.  \label{epsid}
\ee
As in the undeformed case, the action is invariant  under the full set
of local Poincar\'e  transformations (\ref{4dtras}), provided we impose
the torsion to be zero upon making the variations.   The expression
(\ref{4dact}) also differs from that of  the undeformed case due to the fact
that it is not a c-number.  On the other hand, it has definite commutation
properties, i.e. each term in the sum has the same commutation relations with
the connection one forms.

The equations of motion obtained from varying the tetrads have the
usual form, i.e.
\be
\; \epsilon_{abcd} {\cal R}^{ab} \wedge e^{c} = 0 \;,  \label{ffeq1}
\ee
while varying $\omega^{ab}$ gives
\be
\; \epsilon_{abcd} {\cal T}^{c} \wedge e^{d}\; q^{-\Delta(d)} = 0 \;.
\label{ffeq2}
\ee

To make a connection with Einstein gravity, we need to introduce the
space-time metric ${\tt g}_{\mu \nu}$ on $M$. As in the undeformed case, it
has to be a real bilinear in the tetrads which is symmetric in the space-time
indices and invariant under local Lorentz transformations. It should also
reduce to the usual expression in the limit $q\rightarrow 1$. These
requirements uniquely fix ${\tt g}_{\mu\nu}$ to be
\be
{\tt g}_{\mu \nu }=q^{\Delta (a)}\;\eta _{ab}\;e_\mu ^ae_\nu ^b\;,
\label{4dsymmet}
\ee
Using eqs.\eqn{4dcrstc} we see that ${\tt g}_{\mu \nu }$ is symmetric,
although the tensor elements are not c-numbers since
\beqa
{\tt g}_{\mu \nu }\;\omega _\rho ^{ab}=q^{2\Delta (a)+2\Delta (b)}\;
\omega_\rho ^{ab}\;{\tt g}_{\mu \nu }\;,\cr {\tt g}_{\mu \nu }\;e_\rho
^a=q^{2\Delta (a)}\;e_\rho ^a\;{\tt g}_{\mu \nu }\;.
\eeqa
The components of ${\tt g}_{\mu \nu }$ do however commute with themselves.
Finally, one can check that the ${\tt g}_{\mu \nu }$ are real, for both
hermiticity structures i) and ii) given above.

The inverse $e^{\mu}_a$ of the tetrads $e^a_{\mu}$ can be defined
if we enlarge our algebra by a new element ${\tt e}^{-1}$ such that:
\beqa
{\tt e}^{-1} e^a_{\mu}&=& q^{-4 \Delta(a)}e^a_{\mu} \; {\tt e}^{-1}~,\\
{\tt e}^{-1} \omega^{ab}_{\mu}&=& q^{-4 (\Delta(a)+ \Delta(b))
}\omega^{ab}_{\mu}\; {\tt e}^{-1}~,\label{creminus}\eeqa
and $
{\tt e}^{-1} {\tt e} =1$.
Here ${\tt e}$ is the determinant:
$
{\tt e}=\epsilon^{\mu \nu \rho \sigma} e^1_{\mu} e^2_{\nu} e^3_{\rho}
e^4_{\sigma}~,
$
which is consistent because its left hand side  commutes with
everything, due to eqs.\eqn{creminus}. Moreover, one can check that
${\tt e}^{-1} {\tt e} = {\tt e } {\tt e}^{-1}$. The inverse of the
tetrads can now be written:
\be
e^{\mu}_a= \frac{1}{3!} \hat{\epsilon}_{abcd} \epsilon^{\mu \nu \rho
\sigma} e^b_{\nu} e^c_{\rho} e^d_{\sigma} {\tt e}^{-1}~, \label{definv}
\ee
where the totally q-antisymmetric tensor  $\hat{\epsilon}_{abcd} $  is
defined such that
\be
\hat{\epsilon}_{abcd} e^a \wedge e^b \wedge e^c \wedge e^d = e^1
\wedge e^2
\wedge e^3 \wedge e^4 ~~~{\rm no~sum~on}~a,b,c,d
\ee
The solution to this equation can be expressed by
\be
\hat{\epsilon}_{abcd} = q^{3\Delta(a)+2\Delta(b)+\Delta(c) +3}\;
\epsilon_{abcd}  \;.\label{ephaep}\end{equation}
It is easy to prove that the inverse of the tetrads
\eqn{definv} have the usual properties:
\be
e^a_{\mu} e_b^{\mu}=e_b^{\mu} e^a_{\mu} = \delta^a_b~,\;\;\;\;\;\;\;\;
e^a_{\mu} e_a^{\nu}
= e_a^{\nu} e^a_{\mu} = \delta^{\nu}_{\mu}~.
\label{prode}
\ee

By using the inverse of the tetrads, one can now prove that eq.\eqn{ffeq2}
implies the vanishing of the torsion.  Details can be found in
\cite{noi3+1}.

The Christoffel symbols $\Gamma _{\mu \nu }^\sigma $ are defined in the
same way as in the undeformed case, i.e. by
demanding that the covariant derivative of the tetrads vanishes,
\be
0={\tt D}_\mu e_\nu ^b=   \partial _\mu e_\nu ^b+\omega _\mu ^{bc}e_{\nu c}
-\Gamma _{\mu \nu }^\sigma e_\sigma ^b~.\label{covd}
\ee
The difference with the undeformed case is that we cannot switch the
order of the fields arbitrarily. To eliminate the spin--connection from
(\ref{covd}) we now multiply on the left by $q^{\Delta (a)}\eta
_{ab}e_\rho ^a$,  and proceed as in the undeformed case. We can then
isolate $\Gamma _{\mu \nu}^\sigma $ according to
\be
2q^{\Delta (a)}\eta _{ab}e_\rho ^ae_\sigma ^b\Gamma _{\mu \nu }^\sigma
=q^{\Delta (a)}\eta _{ab}[e_\rho ^a(\partial _\mu e_\nu ^b+\partial _\nu
e_\mu ^b)+e_\nu ^a(\partial _\mu e_\rho ^b-\partial _\rho e_\mu ^b)+e_\mu
^a(\partial _\nu e_\rho ^b-\partial _\rho e_\nu ^b)]
\ee
or
\be
2{\tt g}_{\rho \sigma }\Gamma _{\mu \nu }^\sigma =\partial _\mu {\tt g}
_{\rho \nu }+\partial _\nu {\tt g}_{\rho \mu }-\partial _\rho {\tt g}_{\nu
\mu }\;.  \label{qchris}
\ee
To solve this equation we need the inverse of the metric ${\tt g}^{\mu
\nu}$. The expression
\be
{\tt g}^{\mu \nu}= q^{\Delta(a)} \eta^{ab} e^{\mu}_a e^{\nu}_b  \;,
\ee
does the job as it can be checked that
$
{\tt g}^{\mu \rho} {\tt g}_{\rho \nu}=
{\tt g}_{\nu \rho} {\tt g}^{\rho \mu}= \delta^{\mu}_{\nu} \;.
$
Notice that  unlike in the usual Einstein Cartan theory
$
{\tt g}^{\mu \nu} \eta_{ab} e^b_{\nu}= q^{\Delta(a)} e^{\mu}_a~.$
We are now able to solve eq.\eqn{qchris}. Upon multiplying it by ${\tt
g}^{\tau \rho}$ on each side, we get the usual expression for the
Christoffel symbols in terms of the metric tensor and its inverse.  It
may be verified, using these expressions, that the Christoffel symbols
commute with everything and thus, even if written in  terms of
non-commuting quantities, they can be interpreted as being ordinary
numbers.

The covariant derivative operator $\nabla_{\mu}$ defined by the
Christoffel symbols is compatible with the metric ${\tt g}_{\mu \nu}$,
i.e.  $\nabla_{\mu} {\tt g}_{\nu \rho}=0$. This is clear because our
Christoffel symbols have the standard expression in terms of the
space-time metric ${\tt g}_{\mu \nu}$, but it also follows from
eq.\eqn{covd}
\be
\nabla_{\mu} {\tt g}_{\nu \rho}={\tt D}_{\mu} {\tt g}_{\nu \rho}=
{\tt D}_{\mu}( q^{\Delta (a)} \eta_{ab} e^a_{\nu} e^b_{\rho})=0~.
\ee
We now construct the Riemann tensor. It is defined as in the undeformed
theory:
\be
{\tt R}_{\mu \nu \rho}^{~~~~\sigma} v_{\sigma} =({\tt D}_{\mu} {\tt
D}_{\nu}-{\tt D}_{\nu} {\tt D}_{\mu} ) v_{\rho}~,
\ee
where $v_{\mu}$ is a vector. It follows from \eqn{covd}
that it has the standard expression in terms of the Christoffel symbols
(and thus in terms of the space-time metric and its inverse) and
therefore its components commute with everything. (This is also true for
the Ricci tensor ${\tt R}_{\mu \nu}={\tt R}_{\mu \sigma
\nu}^{~~~~\sigma}$, of course, but not for ${\tt R}_{\mu \nu \rho \tau}$
as the lowering of the upper index of the Riemann tensor implies
contraction with ${\tt g}_{\sigma \tau}$ which is not in the center of
the algebra). The relation among the Riemann tensor and the curvature of
the spin connection follows from eq. \eqn{covd}:
\be
e_{\sigma}^a {\tt R}_{\mu \nu \rho}^{~~~~\sigma} v^{\rho}
=e_{\sigma}^a({\tt D}_{\nu} {\tt D}_{\mu}- {\tt D}_{\mu} {\tt D}_{\nu})
 v^{\sigma}=({\cal
D}_{\nu} {\cal D}_{\mu}- {\cal D}_{\mu} {\cal D}_{\nu})e_{\sigma}^a
v^{\sigma}= -{\cal R}_{\mu
\nu }^{ac}\eta_{bc} e_{\sigma}^b v^{\sigma}~,
\ee $ {\cal D}_{\nu}e_{\sigma}^a =  {\partial}_{\nu}e_{\sigma}^a
+  {\omega}_{\nu}^{ab} e_{\sigma b} $, with
  $ {\cal R}^{ab}_{\mu \nu}$ being the space-time components of $ {\cal
R}^{ab}$.  As $v^{\mu}$ is arbitrary, it follows from
the above equation that:
\be
{\tt R}_{\mu \nu \rho}^{~~~~\tau}=-{\cal R}^{ac}_{\mu \nu }\eta_{bc}
e^b_{\rho}e^{\tau}_a~.\label{riecur}\ee
Using this equation it can be checked directly that the components of
the Riemann tensor commute with everything, as pointed out earlier. Our
Riemann tensor has the usual symmetry properties:
\be
{\tt R}_{\mu \nu \rho}^{~~~~\sigma} =  -{\tt R}_{\nu \mu
\rho}^{~~~~\sigma}~,\;\;\;{\tt R}_{\mu \nu
\rho \sigma} = - {\tt R}_{\mu
\nu \sigma \rho}
~,\;\;\;{\tt R}_{[\mu \nu \rho]}^{~~~~~\sigma} =  0~. \label{riem}
\ee
The first of these equations is obvious. The proof of the other two can be
found in Ref. \cite{noi3+1}.

We now show that the action \eqn{4dact} becomes equal to the {\it
undeformed} Einstein-Hilbert action ${\cal S}_{EH}$, once the spin
connection is eliminated using its equations of motion, namely the zero
torsion condition. As in the undeformed case, first we rewrite \eqn{4dact}
in a form analogous to Palatini action, and then show that the latter
reduces to the {\it undeformed} Einstein-Hilbert action, once the
spin-connection is eliminated from it. Consider thus the following
deformation of the Palatini action:
\be
{\cal S}=\frac{1}{2}\int_M d^4 x \, q^{\Delta(a)-3} {\tt e} \, e^{\mu}_a
e^{\nu}_b {\cal R}^{ab}_{\mu \nu}~.
\label{qpala}
\ee
To see that it coincides with \eqn{4dact}, we use the identity:
\be
q^{\Delta(a)-\Delta(b)-6}    \hat{\epsilon}^{abcd}
 e^\mu_a e^\nu_b {\tt e}   =
-\epsilon^{\mu\nu\lambda\sigma} e^c_\lambda e^d_\sigma \;. \label{id68}
\ee
The result (\ref{qpala}) then follows after multiplying both sides of
this equation on the left by
$
-1/8\; q^{-2\Delta(f) - \Delta(g)-3}
\hat{\epsilon}_{fgcd} {\cal R}^{fg}_{\mu\nu}
$
and using the identity
$$
\hat{\epsilon}_{fgcd}
 \hat{\epsilon}^{abcd} =-2 q^6\;(\delta^a_f\delta^b_g-q^
 {\Delta(f)-\Delta(g)}
\delta^a_g\delta^b_f)\;,
$$
along with (\ref{ephaep}).

To show that eq.\eqn{qpala}  in turn is equal to the undeformed
Einstein-Hilbert action, we eliminate the spin connection via its
equation of motion. This amounts to expressing ${\cal R}^{ab}_{\mu \nu}$ in
terms of the Riemann tensor by inverting eq.\eqn{riecur} and then plugging
the result in eq.\eqn{qpala}. We have:
\beqa
&&q^{\Delta(a)} e^{\mu}_a e^{\nu}_b {\cal R}^{ab}_{\mu \nu}=-q^{\Delta(a)}
{\tt R}_{\mu \nu \rho}^{~~~~\tau} e^{\mu}_a e^{\nu}_b e^a_{\tau}
e^{b\rho}=\nonumber\\ &&=-q^{\Delta(b)} {\tt R}_{\mu \nu
\rho}^{~~~~\mu}e^{\nu}_b e^{b\rho}= {\tt R}_{\nu \mu \rho}^{~~~~\mu} {\tt
 g}^{\nu
\rho}={\tt R}~,\label{einhil1}
\eeqa
where we have made use of \eqn{prode}. Moreover we get, after a short
calculation \cite{noi3+1}:
\be
{\tt g} \equiv \det \parallel {\tt g}_{\mu \nu} \parallel =
\frac{1}{4!}\epsilon^{\mu_1 \mu_2 \mu_3
\mu_4} \epsilon^{\nu_1 \nu_2 \nu_3 \nu_4} {\tt g}_{\mu_1 \nu_1}
 {\tt g}_{\mu_2 \nu_2} {\tt g}_{\mu_3 \nu_3} {\tt g}_{\mu_4 \nu_4}
 =q^{-6} \; {\tt
 e}^2~,\label{einhil2}
 \ee
Putting together \eqn{einhil1} and \eqn{einhil2} we see that the
q-Palatini action \eqn{qpala} becomes equal to:
\be
{\cal S}_{EH}=\frac{1}{2}\int_M d^4 x \, \sqrt{-{\tt g}}\; {\tt R}~,
\ee
which is the {\it undeformed} Einstein-Hilbert action. It
is obviously real, since ${\tt g}_{\mu
\nu}$ are real. Moreover, since the components of
${\tt g}_{\mu \nu}$ and its inverse all
commute among themselves, it is clear that the equations of motion of the
metric theory will be equal to those of the undeformed Einstein theory in
vacuum. One can obtain the same result starting directly from
eq.\eqn{ffeq1} and using
\eqn{riecur}.

We now address the issue of the physical interpretation of our construction.
As we have seen above, in our theory the components of the metric tensor
$\tt g_{\mu \nu}$, even though they commute amongst themselves,
are not c-numbers, as
their commutational properties with the
tetrads and the spin-connections are nontrivial. This raises the doubt
that our theory, although formally resembling ordinary general relativity,
may in fact not be {\it physically} equivalent to ordinary general relativity,
 already at the classical level. We remarked that the Christoffel symbols
$\Gamma^{\mu}_{\nu \rho}$ are  c-numbers and, consequently, so is the
Riemann tensor, which encodes most of the geometric information on the
space-time manifold.  Moreover, we can write  the geodesic
equation for a test particle moving in
a gravitational field, which in its parametric form only involves  the
Christoffel symbols:
$
\frac{d^2 x^{\mu}}{d \sigma^2} +\Gamma^{\mu}_{\nu \rho}\frac{d x^{\nu}}
{d \sigma}
\frac{d x^{\rho}}{d \sigma}=F^\mu(\sigma) \;,$
where we have included an arbitrary force $ F^\mu(\sigma)$
 which transforms under diffeomorphisms
like $\frac{d x^{\mu}}{d\sigma}$.

But is this enough to conclude the equivalence of our theory with the
standard metric theory?
After all, the invariant ``distance" $l$ between any two points of
space-time, measured along some path ${\cal C}$ connecting them:
\be
l_{\cal C}=\int_{\cal C} ds
\ee
is an observable quantity.  However, $ds$ constructed from our metric tensor
 is not a
c-number, and hence its physical meaning is unclear.  A    closer
inspection is therefore necessary.   In this regard, we first remark
that when  we
measure a dimensionful quantity (like  length) what we are actually doing
 is  comparing it
with  a standard unit (like a meter).  So we need not require that all
dimensionful quantities be c-numbers.  Rather, what really matters is that the
\underline{ratio} of  any two
quantities carrying the same units is a c-number.
  As we shall see below, this is indeed possible in our description of gravity.

To be definite, let us consider a self-gravitating system of spinless
(electrically) charged point-particles moving in a four dimensional
space-time. In the metric formulation of gravity it is described by the
action (we take $c=1$):
\be
S= \frac{1}{4 \pi G} {\cal S}_{EH} -
\sum_{\alpha}\left\{ m_{\alpha} \int ds^{(\alpha)} + \tilde e_{\alpha}\int
A_{\mu}(x^{(\alpha)})dx^{(\alpha)\mu} \right\} - \frac{1}{16 \pi}\int d^4 x
\sqrt{-\tt g} F_{\mu \nu} F^{\mu \nu}~,\label{act}
\ee
where $G$ is Newton's constant, $m_{\alpha}$ are the masses of the
particles, $A_{\mu}$ is the potential associated with the electromagnetic
 (e.m.) field strength
$F_{\mu\nu}$  and $\tilde e_{\alpha}$ are the electric charges
of the
particles. Using the parameters of this model, we can construct a number of
quantities having the dimension of a length, which play the r\^ole of
natural units for this system. For example, we could define:
\be
l^{(1)}_{\alpha}=G m_{\alpha}~,\;\;\;\;\;\; l^{(2)}_{\alpha}=\frac{\tilde
e^2_{\alpha}}{m_{\alpha}}~.\label{l2}
\ee
If we also use  Planck's constant, we can of course  introduce  the
Planck's length:
\be
l^{(P)}=\sqrt{G \hbar}~.\label{l3}
\ee
 What we shall prove below is that in order for the equations of motions
for the particles and fields to be consistent, the
coupling constants in the action (\ref{act})
 cannot  in general be taken as c-numbers.
(In general, the classical fields need not be c-numbers either.)
Although,  at first sight, this appears to be  a problem, it is in fact
 a blessing, because  the different length units shown above  acquire just
the correct commutation properties needed to make the ratios $l_{\cal
C}/l^{(i)}_{\alpha}$ c-numbers.

In order to see this, let us go back to the action \eqn{act}.  The integral
in the  first term, i.e. the Einstein-Hilbert action ${\cal S}_{EH}$, is
not a c-number.  Even though
 it commutes with ${\tt g}_{\mu
\nu}$, it fails to commute with the tetrads and the spin-connections,
since:
\be
{\cal S}_{EH}\;
 \omega^{ab}_{\mu}= q^{2\Delta(a) +2 \Delta(b)}\omega^{ab}_{\mu}\;
 {\cal S}_{EH}
~,\;\;\;\;\;\; {\cal S}_{EH}\;
 e^{a}_{\mu}=
 q^{2\Delta(a)}e^{a}_{\mu}\;{\cal S}_{EH}\;
 ~.\label{eh}
\ee
Similarly, the integral in the second term,
 giving the interaction between the particles
and the gravitational field, fails to commute
with the tetrads and the spin connections. On the contrary, at least in
four dimensions, we can consistently keep the e.m. potential and the
electric charges as c-numbers.

If we now insist that the action  be a c-number, which seems
desirable if one is to quantize our system using
path-integral techniques, we are forced to give $G$ and
$m_{\alpha}$ nontrivial commutation relations. (Alternatively, if the action
were
not a c-number,  $\hbar$ could not be a c-number.)
 It is easy to verify that
the following commutators  do the job:
$$
G\;\omega^{ab}_{\mu}= q^{2(\Delta(a) +\Delta(b))}\omega^{ab}_{\mu}\;
G~,\;\;\;\;\; G\; e^{a}_{\mu}= q^{2 \Delta(a)}e^{a}_{\mu}\; G~,
\qquad
G \; m_{\alpha} = m_{\alpha} \; G~,$$
\be
m_{\alpha} \;\omega^{ab}_{\mu}= q^{-\Delta(a) -
\Delta(b)}\omega^{ab}_{\mu}\; m_{\alpha}~,\;\;\;\;\;\;\;
m_{\alpha}\; e^{a}_{\mu}= q^{- \Delta(a)}e^{a}_{\mu}\;
m_{\alpha}~,\label{g}
\ee
while the electric charges and fields commute with everything.
We remark that this choice guarantees that $G$ and $m_{\alpha}$ commute
with $\tt g_{\mu\nu}$, so that when deriving the equations of motion from
\eqn{act}, no ordering problem is encountered and one gets the equations of
motion of Einstein gravity.
 The above commutation relations ensure
that the equations of motion are consistent.
Here for simplicity, let us set  all the
electric charges to zero and write down Einstein's equations:
\be
{\tt R}_{\mu\nu}-\frac{1}{2} {\tt g}_{\mu \nu} {\tt R} = 8 \pi G \;
T^{(part)}_{\mu \nu}~,\label{ein}
\ee
where
\be
T^{(part)\mu \nu}(x^0, \bar x) = \sum_{\alpha}
\frac{m_{\alpha}}{\sqrt{-\tt g}}\frac{d x^{(\alpha)\mu}}{d\tau}
\frac{d x^{(\alpha)\nu}}{d x^0} \delta^{(3)}(\bar x - \bar x^{(\alpha)}(x^0))~.
\ee
In the above equation $\bar x$ denotes the ``space''-coordinates $x^i$ and
we think of the particle trajectories $\bar x^{(\alpha)}(x^0)$ as
parametrized by the ``time'' coordinate $x^0$. Thus
$$
\frac{d}{d \tau}= \left\{ {\tt g}_{\mu \nu} \frac{d x^{(\alpha)\mu}}{d x^0}
\frac{d x^{(\alpha)\mu}}{d x^0} \right\}^{-1/2} \frac{d}{d x^0}~.
$$
We see that if the masses and Newton's constant had been c-numbers the
right hand side of \eqn{ein} would not have been a c-number (this can be
 seen commuting it
with, say the tetrads), contrary to the left hand side  of \eqn{ein}.
On the other hand, using the commutation
properties (\ref{g}) solves this difficulty.  Analogous results
hold  for all the equations of motion.

Moreover, one can easily check that the reference lengths
eqs.(\ref{l2}-\ref{l3}) have commutation properties such that the ratios
$l_{\cal C}/l^{(i)}_{\alpha}$ commute with everything:
\be
\left(\frac{l_{\cal C}}{l^{(i)}_{\alpha}}\right)\;\omega^{ab}_{\mu}=
\omega^{ab}_{\mu}\;\left(\frac{l_{\cal C}}{l^{(i)}_{\alpha}}\right)~,
\qquad
\left(\frac{l_{\cal C}}{l^{(i)}_{\alpha}}\right)\;e^{a}_{\mu}=
e^{a}_{\mu}\;\left(\frac{l_{\cal C}}{l^{(i)}_{\alpha}}\right)~,
\ee
and thus are c-numbers, as promised earlier.  (Here we assume the existence of
inverse lengths $1/l^{(i)}_{\alpha}$.)  The same is true, of course,
for the ratio of any two masses. What about the strength of the
gravitational interaction? In this regard, the meaningful thing to do
is to compare the
gravitational force between any two particles with the electric force
between them and thus consider the dimensionless ratios:
$
\frac{G m_{\alpha} m_{\beta}}{\tilde e_{\alpha} \tilde e_{\beta}}~.
$
It easily follows from eqs.(\ref{g}) that this quantity commutes
with everything.

We point out that these conclusions do not depend on the combination of the
parameters that one chooses to construct the unit of measure for the
lengths or the strength of the gravitational interaction. Consider for
example the former. Keeping into account the units of $G$,
$\tilde{e}_{\alpha}$ and $\hbar$ (we remind the reader that we are assuming
$c=1$ and thus a length has the same dimension as a time), it is easy to
check that a monomial of the form:
\be
{\cal P}=G^{p}m^q \tilde{e}^{2r} \hbar^s
\ee
will have the dimensions of a length if and only if:
\be
p+r+s=1~,\;\;\;\;\;\;\;\;
-p + q +r+s=0~.\label{len}
\ee
Consider now commuting $\cal P$ with, say, the tetrads. One finds:
\be
{\cal P}\; e^a_{\mu}=q^{(2p-q)\Delta(a)}e^a_{\mu}\;{\cal P}
\ee
Subtracting the second of eqs.\eqn{len} from the first, we see that
$2p-q=1$ and thus the above commutator can be rewritten as:
\be
{\cal P}\; e^a_{\mu}=q^{\Delta(a)}e^a_{\mu}\;{\cal P}~.
\ee
Repeating the same computation for the commutator of ${\cal P}$ with the
spin-connections would give:
\be
{\cal P}\;
\omega^{ab}_{\mu}=q^{\Delta(a)+\Delta(b)}\omega^{ab}_{\mu}\;{\cal P}~.
\ee
Thus the ratios $l_{\cal C}/{\cal P}$ are c-numbers for any choice of
${\cal P}$. An analogous result holds for any monomial ${\cal P}$ with the
units of $G$.

 In view of the above considerations, we can now claim the
complete
\underline{physical} equivalence of our theory with Einstein gravity.

From the above results  we may conclude that if we just consider the
theory constructed in terms of the space-time metric ${\tt g}_{\mu \nu}$,
ignoring the underlying gauge formulation,  our theory is completely
equivalent to Einstein's theory.  No trace of the non--commutative
structure existing in the gauge formulation of the theory can be found at
the metric level. Though the metric  does not commute with the connection
components, all the physical objects constructed out of it, e. g. the
Christoffel symbols together with the Riemann, Ricci and Einstein tensors,
are c-numbers. Thus it appears that, at the level of {\it classical} general
relativity we can choose whatever representative of the one parameter
family of $q$--gauge theories (not only the well known $q=1$ theory)
without changing the physics we are describing. That is, we have discovered
a non--commutative structure in general relativity which is hidden, even in
the presence of matter, provided there are no sources of torsion.

{\bf Acknowledgments}

R.M. acknowledges E.E.C. contract ERBFMRX-CT96-0045 and a grant of the Italian
Ministry of Education and Scientific Research.  A.S. was supported in part by
the U.S. Department of Energy under contract  number DE-FG05-84ER40141.
A.S. would like to thank the members of the theory group at the University of
Naples for their warm hospitality while this work was completed.

\end{document}